\documentclass[11pt,a4paper,twoside,groupcitations]{article}
\usepackage[T1]{fontenc}
\usepackage[ansinew]{inputenc}
\usepackage[english]{babel}
\usepackage{amsfonts}
\usepackage{amsmath}
\usepackage{bm}
\usepackage{array}
\usepackage{amsthm}
\usepackage{amssymb}
\usepackage{graphicx}
\usepackage{subfigure}
\usepackage{braket}
\usepackage{eucal}
\usepackage{verbatim}
\usepackage[table]{xcolor}
\usepackage{caption}
\usepackage{cite}
\usepackage{textcomp}
\usepackage{multicol}
\usepackage{tikz}
\usetikzlibrary{positioning,arrows}
\usetikzlibrary{decorations.pathmorphing}
\usetikzlibrary{decorations.markings}
\usetikzlibrary{calc,decorations.markings}
\usetikzlibrary{arrows,shapes}
\usetikzlibrary{matrix,arrows}
\usepackage{pgfplots}
\usepackage{xparse}
\definecolor{jade}{HTML}{00A86B}
\newcommand{\be}{\begin{eqnarray}}
\newcommand{\ee}{\end{eqnarray}}
\newcommand{\pro}[2]{\mbox{$\langle\, #1 \mid #2\,\rangle$}}

\renewcommand{\d}{\mbox{${\rm d}$}} 
\newcommand{\lp}{\ell_{\rm p}}
\newcommand{\mpl}{m_{\rm p}}
\newcommand{\gn}{G_{\rm N}}
\newcommand{\rh}{r_{\rm H}}

\title{\bf Quantum dust cores of rotating black holes}
\author{Tommaso~Bambagiotti$^{ab}$\thanks{E-mail: tommaso.bambagiotti2@unibo.it}
$\,$
and
Roberto~Casadio$^{abc}$\thanks{E-mail: casadio@bo.infn.it}
\\
\\
$^a${\em Dipartimento di Fisica e Astronomia, Universit\`a di Bologna}
\\
{\em via Irnerio~46, 40126 Bologna, Italy}
\\
\\
$^b${\em I.N.F.N., Sezione di Bologna, I.S.~FLAG}
\\
{\em viale B.~Pichat~6/2, 40127 Bologna, Italy}
\\
\\
$^c${\em Alma Mater Research Center on Applied Mathematics (AM$^2$)}
\\
{\em Via Saragozza 8, 40123 Bologna, Italy}
}
\begin{document}
\maketitle
\begin{abstract}
Black holes are spacetimes that should describe the end state of the gravitational
collapse of huge amounts of quantum matter. 
A quantum description of dust cores for black hole geometries that accounts for the large number
of matter constituents can be obtained by quantising the geodesic motion of dust particles and finding
the corresponding many-body ground state.
We here generalise previous works in spherical symmetry to rotating geometries and show the effect of
angular momentum on the size of the core and effective interior geometry.
\end{abstract}
\section{Introduction}
\setcounter{equation}{0}
\label{S:intro}
Astrophysical black holes~\cite{Fabian:2019sxb} are expected to result from the gravitational collapse
of huge amounts of matter which, in turn, can be properly described only by quantum physics.
Understanding the nature of such objects therefore requires us to take into account both the complexity
and the quantum features of their matter source.
Disregarding the quantum nature of matter generically leads to end states described by singular
spacetimes~\cite{Penrose:1964wq,Senovilla:2014gza,HE}. 
For example, the Oppenheimer-Snyder model of homogenous dust balls~\cite{Oppenheimer:1939ue}
ends in the vacuum Schwarzschild solution~\cite{Schwarzschild:1916uq} of the Einstein equations.
\par
Classical spacetime singularities can be removed by imposing regularity conditions on the (effective)
energy density and scalar invariants inspired by classical physics~\cite{Carballo-Rubio:2023mvr}.
This procedure usually induces the appearance (or fails to remove) an inner Cauchy horizon. 
A quantum framework, invoked for example in Ref.~\cite{Casadio:2023iqt}, can be implemented
by describing the collapsed matter inside black holes with an effective energy density $\rho\propto|\psi|^2$,
where $\psi$ is the matter wavefunction in the Madelung approximation~\cite{Madelung:1927ksh}.
Normalisability of $\psi=\psi(r)$ then implies that the Misner-Sharp-Hernandez (MSH) mass
function~\cite{Misner:1964je,Hernandez:1966zia} satisfies
\be
m(r)
\equiv
4\,\pi
\int_0^r
\rho(x)\,x^2\,\d x
\sim
4\,\pi
\int_0^r
|\psi(x)|^2\,x^2\,\d x
<
\infty
\qquad
{\rm for}
\ r>0
\ .
\label{Qcond}
\ee
This accommodates for $\rho\sim r^{-2}$ and $m\sim r$, which ensures that $m(0)=0$
and replaces the central singularity with an integrable singularity~\cite{Casadio:2021eio,Casadio:2022ndh},
that is a region where the curvature invariants and the effective energy density and pressures diverge
but their volume integrals remain finite~\cite{Lukash:2013ts}.
An additional feature of relevance is that the interior does not contain Cauchy horizons.
\par
Most attempts at quantising models of the gravitational collapse start from a reduced Einstein-Hilbert action 
in order to define a mini-superspace~\cite{DeWitt:1967yk,kiefer} for a very small number of
degrees of freedom.
For example, the canonical quantisation of the Oppenheimer-Snyder model~\cite{Oppenheimer:1939ue}
employed in Refs.~\cite{Vaz:2011zz,Kiefer:2019csi,Husain:2022gwp,Giesel:2022rxi}, usually
results in a wavefunction for the radius and Arnowitt-Deser-Misner (ADM) mass~\cite{Arnowitt:1959ah}
of the dust ball.
One then typically finds that the semiclassical evolution displays a bounce from collapsing
to expanding radius~\cite{Casadio:1998yr}.
Since the radius and total mass of an astrophysical object are collective degrees of freedom,
their quantisation is tantamount to introducing an uncertainty relation for the thermodynamic variables
of a macroscopic fluid, and clearly ignores the complexity inherent in the collapse of huge amounts
of matter into self-gravitating bound states.
\par
To account for the many-body nature of the collapsed matter in Ref.~\cite{Casadio:2023ymt},
the dust ball was described as a sequence of layers~\cite{Tolman:1934za} of particles,
whose trajectories were individually quantised (see also Refs.~\cite{Casadio:2021cbv,Casadio:2022pla}).
A condition was then imposed to ensure that the quantum layers did not cross in the global quantum
ground state, which leads to a dust core of macroscopic size.
We stress that this approach does not involve quantising the background geometry or its perturbations
in any ways.
In a complementary perspective, dust particles are viewed as the fundamental physical constituents
of the bound states they form under their own gravitational pull.
Since the ground state should not evolve any further (unless it is perturbed from the outside or
the Hawking effect~\cite{Hawking:1974rv} is included), it remains an open question whether
the dust core admits a description in terms of a Lorentzian metric for the black
hole interior.~\footnote{One might instead consider a Carrollian metric~\cite{Ciambelli:2025unn,Ecker:2023uwm}.}
In fact, one might view such a quantum dust core as the state with no classical counterpart
in the tunnelling phase of the black hole-to-white hole transition~\cite{Haggard:2014rza} that
leads to the semiclassical bounce mentioned above.
Nonetheless, one can compute an effective metric for the dust core which yields an effective energy density
and MSH mass function satisfying the condition in Eq.~\eqref{Qcond}~\cite{Casadio:2023ymt,Gallerani:2025wjc}.
\par
In this work, we consider (differentially) rotating cores of dust by explicitly employing geodesic motion
in the (generalised~\cite{Casadio:2023iqt}) Kerr metric~\cite{Kerr:1963ud} and compare with the perturbative
results of Ref.~\cite{Casadio:2022epj}, where angular momentum was added to the spherical symmetric geometry.
We will find that the fully general relativistic treatment of rotating dust leads to ground states representing
elongated cores of smaller size compared to the spherically symmetric case.
We will also identify a ground state configuration in which the effective mass function and specific
angular momentum grow linearly inside the dust core, as expected for an integrable singularity
without Cauchy horizons of the form discussed in Ref.~\cite{Casadio:2023iqt}.
\section{Rotating geodesics}
\setcounter{equation}{0}
\label{S:geodesics}
We will describe the collapsing matter as dust particles moving in a generalised Kerr geometry
which, in Boyer-Lidquist coordinates $x^\alpha=(t,r,\theta,\phi)$, reads~\footnote{We shall always use
units with $c=1$ and often write the Planck constant $\hbar=\lp\,\mpl$ and the Newton constant
$\gn=\lp/\mpl$, where $\lp$ and $\mpl$ are the Planck length and mass, respectively.}
\be
\d s^{2}
=
-\d t^2
+
\frac{2\,\gn\,m\,r}{\rho^2}
\left(a\,\sin^2\theta\,\d\phi-\d t\right)^2
+
\rho^2
\left(
\frac{\d r^2}{\Delta}
+\d\theta^{2}
\right)
+
\left(r^2+a^2\right)
\sin^2\theta\,\d\phi^2
\ ,
\label{kerr}
\ee
where
\be
\rho^2
=
r^2+a^2\,\cos^2\theta
\ee
and
\be
\Delta
=
r^2-2\,\gn\,m\,r+a^2
\ .
\label{e:delta}
\ee
In the above expressions, the function $m=m(r)$ represents the MSH mass inside ellipsoids of coordinate
radius $r$ and $a=a(r)=J(r)/m(r)$ is the specific angular momentum on the surface of the same
ellipsoid~\cite{Casadio:2023iqt}. 
\par
We recall that the vacuum Kerr metric~\cite{Kerr:1963ud} is given by constant $a=A$ and $m=M$,
where $M$ is now the Arnowitt-Deser-Misner (ADM) mass~\cite{Arnowitt:1959ah},
and it contains horizons located at
\be
\label{eq:horRad}
R_\pm
=
\gn\,M
\pm
\sqrt{\gn^2\,M^2-A^2}
\ ,
\label{R+-}
\ee
provided $A^2\le \gn^2\,M^2$.
We expect $m$ and $a$ to approach asymptotically $M$ and $A$, respectively, in the vacuum outside
the collapsing dust.
\subsection{Action and equations of motion}
Assuming individual dust particles have a proper mass $\mu\ll m$, their trajectories
can be approximated by time-like geodesics $x^\alpha=x^\alpha(\tau)$ in the metric~\eqref{kerr},
governed by the Lagrangian
\be
2\,L
&\!\!=\!\!&
\dot t^2
-
\frac{2\,\gn\,m\,r}{\rho^2}
\left(a\,\sin^2\theta\,\dot\phi-\dot t\right)^2
-
\rho^2
\left(
\frac{\dot r^2}{\Delta}
+\dot\theta^{2}
\right)
-
\left(r^2+a^2\right)
\sin^2\theta\,\dot\phi^2
\ ,
\label{kerrL}
\ee
which yields the mass-shell condition $2\,L=1$ and the integrals of motion
\be
\frac{E}{\mu}
=
\left(1-\frac{2\,\gn\,m\,r}{\rho^2}\right)
\dot t
+
\frac{2\,\gn\,m\,a\,r}{\rho^2}\,
\sin^2\theta\,\dot\phi
\ee
and
\be
j
=
\left(
r^2+a^2+\frac{2\,\gn\,m\,a^2\,r}{\rho^2}\,\sin^2\theta
\right)
\sin^2\theta\,\dot\phi
-
\frac{2\,\gn\,m\,a\,r}{\rho^2}\,
\sin^2\theta\,\dot t
\ .
\ee
We can invert the above relations and obtain
\be
\dot t
&\!\!=\!\!&
\frac{1}{\Delta}
\left[
\left(
r^2+a^2+\frac{2\,\gn\,m\,a^2\,r}{\rho^2}\,\sin^2\theta
\right)
\frac{E}{\mu}
-
\frac{2\,\gn\,m\,a\,r}{\rho^2}\,j
\right]
\nonumber
\\
&\!\!=\!\!&
\frac{E}{\mu}
+
\frac{4\,\gn\,m\,r\left[(r^2+a^2)\,E/\mu+a\,j\right]}
{\left(r^2+a^2-2\,\gn\,m\,r\right)\left[2\,r^2+a^2\left(1+\cos2\,\theta\right)\right]}
\ee
and
\be
\dot\phi
&\!\!=\!\!&
\frac{1}{\Delta}
\left[
\left(\frac{2\,\gn\,m\,a\,r}{\rho^2}
\right)
\frac{E}{\mu}
+
\left(
1
-
\frac{2\,\gn\,m\,r}{\rho^2}
\right)
\frac{j}{\sin^2\theta}
\right]
\nonumber
\\
&\!\!=\!\!&
\frac{4\,\gn\,m\,a\,r\,E/\mu+2\,a^2\,j}
{\left(r^2+a^2-2\,\gn\,m\,r\right)\left[2\,r^2+a^2\left(1+\cos2\,\theta\right)\right]}
-
\frac{2\,j}{\left[2\,r^2+a^2\left(1+\cos2\,\theta\right)\right]\,\sin^2\theta}
\ .
\ee
\par
We next assume that the dust particles in a layer at the surface of an ellipsoid of radial coordinate
$r=r(\tau)$ co-rotate with the geometry and therefore have $j=0$, that is
\be
\left(
r^2+a^2+\frac{2\,\gn\,m\,a^2\,r}{\rho^2}\,\sin^2\theta
\right)
\dot\phi
=
\frac{2\,\gn\,m\,a\,r}{\rho}\,
\dot t
\ ,
\ee
which implies
\be
\dot t
&\!\!=\!\!&
\left[
1
+
\frac{2\,\gn\,m\,r\,(r^2+a^2)}{\Delta\,\rho^2}
\right]
\frac{E}{\mu}
\nonumber
\\
&\!\!=\!\!&
\left\{
1
+
\frac{4\,\gn\,m\,r\,(r^2+a^2)}
{\left(r^2+a^2-2\,\gn\,m\,r\right)\left[2\,r^2+a^2\left(1+\cos2\,\theta\right)\right]}
\right\}
\frac{E}{\mu}
\ee
and
\be
\dot\phi
&\!\!=\!\!&
\left(\frac{2\,\gn\,m\,a\,r}{\Delta\,\rho^2}\right)
\frac{E}{\mu}
\nonumber
\\
&\!\!=\!\!&
\left\{
\frac{4\,\gn\,m\,a\,r}
{\left(r^2+a^2-2\,\gn\,m\,r\right)\left[2\,r^2+a^2\left(1+\cos2\,\theta\right)\right]}
\right\}
\frac{E}{\mu}
\ .
\ee
With this assumption, the Lagrangian~\eqref{kerrL} simplifies to
\be
2\,L_0
&\!\!=\!\!&
-\rho^2\left(\frac{\dot r^2}{\Delta}+\dot\theta^2\right)
+
\left[
1
+
\frac{2\,\gn\,m\,r\,(r^2+a^2)}{\Delta\,\rho^2}\,
\right]^2
\frac{E^2}{\mu^2}
-
\left[
\frac{2\,\gn\,m\,r\,(r^2+a^2)^{2}}{\Delta^2\,\rho^2}
\right]
\frac{E^2}{\mu^2}
\nonumber
\\
&&
-
\left[
\frac{4\,\gn^2\,m^2\,a^2\,r^2\,(r^2+a^2)\,\sin^2\theta}{\Delta^2\,\rho^4}
\right]
\frac{E^2}{\mu^2}
=
1
\ .
\label{kerrL0}
\ee
We note that, for $E=0$, the above expression greatly simplifies to
\be
2\,L_0
=
-\rho^2\left(\frac{\dot r^2}{\Delta}+\dot\theta^2\right)
=
1
\ ,
\ee
which can be written as
\be
\frac{1}{2}\,\dot r^2
+
\frac{\Delta}{2\,\rho^2}
+
\frac{\Delta}{2}\,\dot\theta^2
=
\frac{1}{2}\,\dot r^2
+
\frac{r^2-2\,\gn\,m\,r+a^2}{2\left(r^2+a^2\,\cos^2\theta\right)}
+
\frac{1}{2}
\left(r^2-2\,\gn\,m\,r+a^2\right)
\dot\theta^2
=
0
\ .
\ee
The equation of motion for $\theta=\theta(\tau)$ then reads
\be
\ddot\theta
=
\frac{1}{2}\,
\frac{\partial}{\partial\theta}
\left(
\frac{1}{r^2+a^2\,\cos^2\theta}
\right)
=
\frac{a^2\,\cos\theta\,\sin\theta}{\left(r^2+a^2\,\cos^2\theta\right)^2}
\ ,
\ee
so that $\theta=\theta(\tau)$ can be constant for $E=j=0$ only if $\theta=0$ or $\theta=\pi/2$.
\par
Let us next consider the case of radial motion along the axis of symmetry $\theta=\dot\theta=0$,
and motion on the equatorial plane $\theta=\pi/2$ and $\dot\theta=0$.
\subsubsection{Axial motion}
For $\theta=\dot\theta=0$, the Lagrangian~\eqref{kerrL0} reads
\be
2\,L_0
&\!\!=\!\!&
\left(1-\frac{2\,\gn\,m\,r}{\rho^2}\right)^{-1}
\frac{E^2}{\mu^2}
-
\frac{\rho^2}{\Delta}\,
\dot r^{2}
=
1
\ ,
\label{kerrL0a}
\ee
which can be written as
\be
\frac{1}{2}\,\mu\,\dot r^2
-
\frac{\gn\,\mu\,m}{r}
\left(
1
-
\frac{a^2}{r^2+a^2}
\right)
=
\frac{\mu}{2}
\left(
\frac{E^2}{\mu^2}
-
1
\right)
\ .
\label{Tax}
\ee
Note that the above equation describes purely radial motion in a Schwarzschild spacetime for $a=0$.
\subsubsection{Equatorial motion}
For $\theta=\pi/2$ and $\dot\theta=0$, the Lagrangian~\eqref{kerrL0} becomes
\be
2\,L_0
&\!\!=\!\!&
\frac{(E^2/\mu^2-\dot r^2)\,r^3+a^2\,(r+2\,\gn\,m)\,E^2/\mu^2}
{r\left(r^2+a^2-2\,\gn\,m\,r\right)}
=
1
\ ,
\label{kerrL0e}
\ee
which can be rewritten as
\be
\frac{1}{2}\,\mu\,\dot r^2
-
\mu\left(
\frac{\gn\,m}{r}
-
\frac{a^2}{2\,r^2}
\right)
-
\left(1+\frac{2\,\gn\,m}{r}\right)
\frac{a^2\,E^2}{2\,\mu\,r^2}
=
\frac{\mu}{2}
\left(
\frac{E^2}{\mu^2}
-
1
\right)
\ .
\label{Teq}
\ee
Note that the last term in the left hand side couples the energy $E$ of the dust particle with the
(specific) angular momentum $a$ of the system.
\section{Ground state and perturbative spectrum}
\setcounter{equation}{0}
\label{S:Spert}
We can discretise the rotating ball by considering an ellipsoidal core of mass $\mu_0=\epsilon_0\,M$
and coordinate radius $r=R_1(\tau)$ surrounded by $N$ comoving layers of inner radius $r=R_i(\tau)$,
thickness $\Delta R_i=R_{i+1}-R_i$, and mass $\mu_i=\epsilon_i\,M$, where $\epsilon_i$ is the fraction
of ADM mass carried by the $\nu_i$ dust particles in the $i^{\rm th}$ layer.
The gravitational mass inside the ellipsoid $r<R_i$ will be denoted by 
\be
M_i
=
\sum_{j=0}^{i-1}
\mu_j
=
M\,\sum_{j=0}^{i-1}\epsilon_j
\ ,
\ee
with $M_1=\mu_0$ and $M_{N+1}=M$. 
Likewise, we denote with $A_i$ the specific angular momentum at the inner surface of the $i^{\rm th}$ layer. 
\par
The evolution of each layer can be derived by noting that dust particles located on the symmetry axis
or on the equator at $r=R_i(\tau)$ will follow the geodesic equation
\be
\label{geod-part}
H_i
\equiv
\frac{P_i^{2}}{2\,\mu}
-\frac{\gn\,\mu\,M_i}{R_i}
+
\mu\,W_i
=
\frac{\mu}{2}\left(\frac{E_i^{2}}{\mu^2}-1\right)
\equiv
\mathcal E_i
\ ,
\ee
where $P_i=\mu\,\d R_i/\d\tau$ is the radial momentum conjugated to $R=R_i(\tau)$, $E_i$
the conserved momentum per unit mass conjugated to $t=t_i(\tau)$ and $W_i$ denotes the other terms
in the radial potential.
In particular, for the axial motion
\be
W_i
=
\frac{\gn\,M_i\,A_i^2}{R_i\left(R_i^2+A_i^2\right)}
\equiv
W_i^{\rm ax}
\ ,
\ee
whereas for equatorial motion
\be
W_i
=
\frac{A_i^2}{2\,R_i^2}
\left[
1
-
\left(
1
+
\frac{2\,\gn\,M_i}{R_i}
\right)
\frac{E_i^2}{\mu^2}
\right]
\equiv
W_i^{\rm eq}
\ .
\ee
With the canonical quantization prescription $P_i\mapsto\hat{P}_i=-i\,\hbar\,\partial_{R_i}$,
Eq.~\eqref{geod-part} becomes the time-independent Schr\"odinger equation
\be
\hat{H}_i\,\psi_{n_i}
=
\left[
-
\frac{\hbar^{2}}{2\,\mu}
\left(
\frac{\d^2}{\d R_i^2}
+
\frac{2}{R_i}\,
\frac{\d}{\d R_i}
\right)
-
\frac{\gn\,\mu\,M_i}{R_i}
+
\mu\,W_i
\right]
\psi_{n_i}
=
\mathcal E_{n_i}\,
\psi_{n_i}
\ ,
\label{SchE}
\ee
where the kinetic term was defined so as to ensure that the quantum states
describe a spatially 3-dimensional system like the original Lagrangian~\eqref{kerrL}.
In other words, the effective Hamiltonian constraint~\eqref{SchE} is obtained by quantising the
dynamical system that describes all possible geodesics and then imposing the condition $j=0$
that leads to Eq.~\eqref{kerrL0}.
We remark that the radius $R$ is a natural choice for the quantisation of dust particles
in a rotating spacetime because it has the invariant geometric meaning of identifying the
ellipsoids that represent the surfaces of symmetry of the system.
Likewise, the proper time along geodesics is a scalar which does not depend on the
choice of any coordinates.
\subsection{Non-rotating case}
When $W_i\sim a^2$ is negligible, Eq.~\eqref{SchE} is analogous to the equation for $s$-states
of the hydrogen atom, and the solutions are given by the eigenfunctions~\cite{Casadio:2023ymt}
\be
\psi_{n_i}(R_i)
=
\sqrt{\frac{\mu^6\,M_i^{3}}{\pi\,\lp^{3}\,\mpl^{9}\,n_i^{5}}}\,
\exp\!\left(-\frac{\mu^2\,M_i\,R_i}{n_i\,\mpl^{3}\,\lp}\right)
L_{n_i-1}^{1}\!\!
\left(\frac{2\,\mu^2\,M_i\,R_i}{n_i\,\mpl^{3}\,\lp}\right)
\ ,
\label{radial-wavefunction}
\ee
where $L_{n-1}^1$ are Laguerre polynomials and ${n}_i=1,2\,\ldots$, corresponding to the eigenvalues 
\be
\mathcal E_{n_i}^{(0)}
=
-
\frac{\mu^3\,M_i^2}{2\,\mpl^4\,n_i^2}
\ .
\ee
The wavefunctions~\eqref{radial-wavefunction} are normalised in the scalar product which makes $\hat H_i$ Hermitian
for $W_i=0$, that is
\be
\pro{n_i}{n'_i}
=
4\,\pi
\int_0^\infty
R_i^2\,\psi_{n_i}^*(R_i)\,\psi_{n'_i}(R_i)\,
\d R_i
=
\delta_{n_i n'_i}
\ .
\ee
The expectation value of the coordinate radius on these eigenstates is given by
\be
\bar R_{n_i}
\equiv
\bra{n_i} \hat R_i \ket{n_i}
=
\frac{3\,\mpl^3\,\lp\,n_i^2}{2\,\mu^2\,M_i}
\ ,
\ee
with relative uncertainty
\be
\frac{\overline{\Delta R}_{n_i}}{\bar R_{n_i}}
\equiv
\frac{\sqrt{\bra{n_i}\hat R^2_i\ket{n_i}-\bar R_{n_i}^2}}{\bar R_{n_i}}
=
\frac{\sqrt{n_i^2+2}}{3\,n_i}
\ ,
\label{DR/R}
\ee
which approaches the minimum $\overline{\Delta R}_{n_i}\simeq \bar R_{n_i}/3$ for $n_i\gg 1$.
\par
By assuming that the conserved quantity $E_i$ remains well-defined for all the dust particles
in the allowed quantum states, we obtain the fundamental condition~\cite{Casadio:2021cbv}
\be
0
\le
\frac{E_i^2}{\mu^2}
=
1
+
\frac{2\,\mathcal E_i^{(0)}}{\mu}
=
1
-
\frac{\mu^2\,M_i^2}{\mpl^4\,n_i^2}
\ ,
\label{Emu}
\ee
which yields the lower bound for the single particle principal quantum numbers
\be
n_i
\ge
N_i^{(0)}
\equiv
\frac{\mu\,M_i}{\mpl^2}
\ .
\label{N_M}
\ee
Upon saturating the above bound, one then finds
\be
\bar R_{N_i}^{(0)}
=
\frac{3}{2}\,\gn\,M_i
\ ,
\label{RNi}
\ee
and the wavefunction for the $\nu_i$ particles in each layer is given by the same ground state
\be
\psi_{N_i}^{(0)}(R_i)
=
\sqrt{\frac{\mu\,\mpl}{\pi\,\lp^{3}\,M_i^{2}}}\,
\exp\!\left(-\frac{\mu\,R_i}{\mpl\,\lp}\right)
L_{\frac{\mu\,M_i}{\mpl^2}-1}^{1}\!\!
\left(\frac{2\,\mu\,R_i}{\mpl\,\lp}\right)
\ ,
\label{groundI}
\ee
where the values of $M_i$, hence $N_i^{(0)}$ in Eq.~\eqref{N_M}, must be such that $\bar R_i^{(0)}\lesssim \bar R_{i+1}^{(0)}$.
\subsection{Slow-rotation corrections}
We can estimate the corrections to the radial potential using the above result for the ground state
$E_i=0$ with $R_i\simeq \bar R_{N_i}^{(0)}$.
For the axial motion, we find
\be
W_i^{\rm ax}
\simeq
\frac{\gn\,M_i\,A_i^2}{R_i\left(R_i^2+A_i^2\right)}
\simeq
\frac{8\,A_i^2}{3\left(9\,\gn^2\,M_i^2+4\,A_i^2\right)}
\simeq
\frac{8\,A_i^2}{27\,\gn^2\,M_i^2}
\ ,
\label{eq:axiPer}
\ee
whereas for equatorial motion
\be
W_i^{\rm eq}
\simeq
\frac{A_i^2}{2\,R_i^2}
\left[
1
-
\left(
1
+
\frac{2\,\gn\,M_i}{R_i}
\right)
\frac{E_i^2}{\mu^2}
\right]
\simeq
\frac{2\,A_i^2}{9\,\gn^2\,M_i^2}
\ .
\label{eq:equPer}
\ee
For perturbation theory to apply, the above potentials must be smaller than
\be
V_i
=
\frac{\gn\,M_i}{R_i}
\simeq
\frac{\gn\,M_i}{3\,\gn\,M_i}
\simeq
\frac{1}{3}
\ .
\label{V_i}
\ee 
This yields
\be
\frac{A_i^2}{\gn^2\,M_i^2}
\ll
\frac{27}{24}
\label{e:axPer}
\ee
for axial motion and 
\be
\frac{A_i^2}{\gn^2\,M_i^2}
\ll
\frac{3}{2}
\label{e:eqPer}
\ee
for equatorial motion.
Both conditions above are approximately satisfied for classical Kerr black holes (with $A^2\le \gn^2\,M^2$).
\par
Since $W_i$ is constant to leading order in $A_i$, it will simply result in a shift of the energy
eigenvalues
\be
\mathcal E_i
=
\mathcal E_i^{(0)}
+
\mu\,W_i
\ .
\ee
For the ground state, we then have
\be
0
=
\frac{E_i^2}{\mu^2}
=
1
+
\frac{2\,\mathcal E_i}{\mu}
=
1
-
\frac{\mu^2\,M_i^2}{\mpl^4\,N_i^2}
+
2\,W_i
\ ,
\label{Emu1}
\ee
or
\be
N_i
\simeq
\frac{\mu\,M_i}{\mpl^2\left(1+2\,W_i\right)^{1/2}}
\simeq
\frac{\mu\,M_i}{\mpl^2}
\left(1-W_i\right)
=
N_i^{(0)}
\left(1-W_i\right)
\ .
\label{NiW}
\ee
Correspondingly, we obtain a reduction in the areal radius
\be
\bar R_{N_i}
=
\frac{3\,\mpl^3\,\lp\,N_i^2}{2\,\mu^2\,M_i}
\simeq
\bar R_{N_i}^{(0)}
\left(1-2\,W_i\right)
\ ,
\label{bRNi}
\ee
with unaffected (leading order) uncertainty.
Note in particular that 
\be
\bar R_{N_i}^{\rm ax}
<
\bar R_{N_i}^{\rm eq}
<
\bar R_{N_i}^{(0)}
\ ,
\label{rAxEq}
\ee
so that the effect of rotation is to reduce the size of the layers.
\par
This behaviour is the opposite of what was obtained in Ref.~\cite{Casadio:2022epj},
where angular momentum was added to the quantum states of dust particles but assuming
the Schwarzschild metric still determines the geodesic motion.
For radial geodesics, the addition of angular momentum therefore amounted to a repulsive
centrifugal term in the radial potential which made the size of the core increase.
In the present, fully general relativistic treatment, the angular momentum contributes
both a repulsive centrifugal term and an attractive term in Eqs.~\eqref{Tax} and \eqref{Teq},
with the latter overcoming the former for $r\lesssim 2\,\gn\,M$.
\subsubsection{Core size}
The above result also agrees with the expected shape of the core being elongated on the equatorial 
plane with respect to the symmetry axis in spherical coordinates, which are the ones used
to express the states~\eqref{groundI}.
For the outermost layer (the surface of the core),  
\be
R_{\rm s}
\simeq
\frac{4}{3}\,\bar R_{N_N}
\simeq
\frac{3}{2}\,\gn\,M
\left(1-2\,W_N\right)
\ ,
\label{RsW}
\ee
along the axis of symmetry we have
\be
R_{\rm s}^{\rm ax}
\simeq
\frac{3}{2}\,\gn\,M
\left(1-\frac{8\,A^2}{27\,\gn^2\,M^2}\right)
\ ,
\ee
where $A$ is the constant specific angular momentum of the outer Kerr geometry.
On the equatorial plane one similarly finds
\be
R_{\rm s}^{\rm eq}
\simeq
\frac{3}{2}\,\gn\,M
\left(1-\frac{2\,A^2}{9\,\gn^2\,M^2}\right)
\ ,
\ee
so that 
\be
\frac{R_{\rm s}^{\rm eq}}{R_{\rm s}^{\rm ax}}
\simeq
1-2\,W_N^{\rm eq}+2\,W_N^{\rm ax}
\simeq
1
+
\frac{2\,A^2}{27\,\gn^2\,M^2}
\ .
\ee
\begin{figure}[t]
\centering
\includegraphics[width=8.5cm]{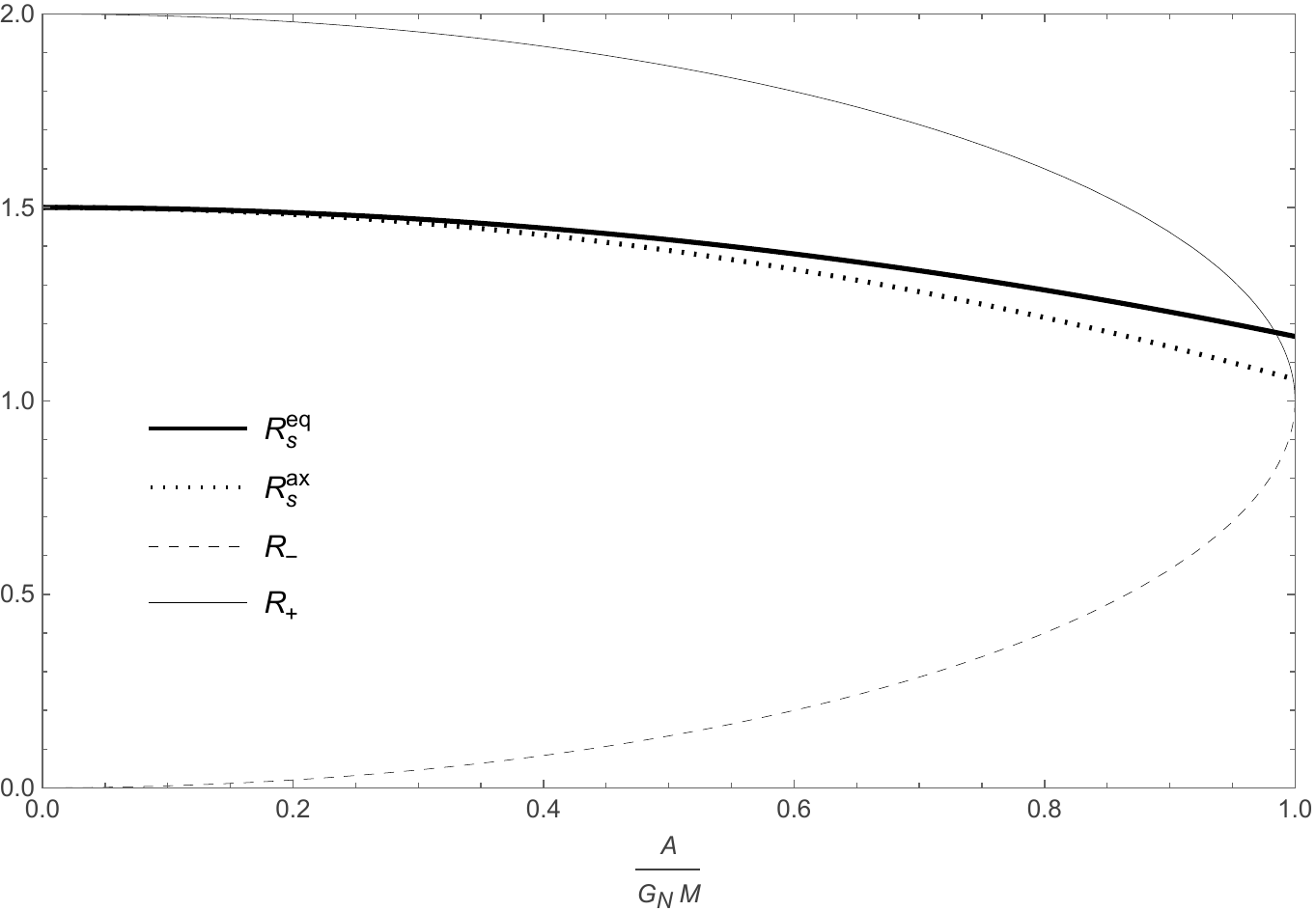}
$\quad$
\includegraphics[width=6.5cm]{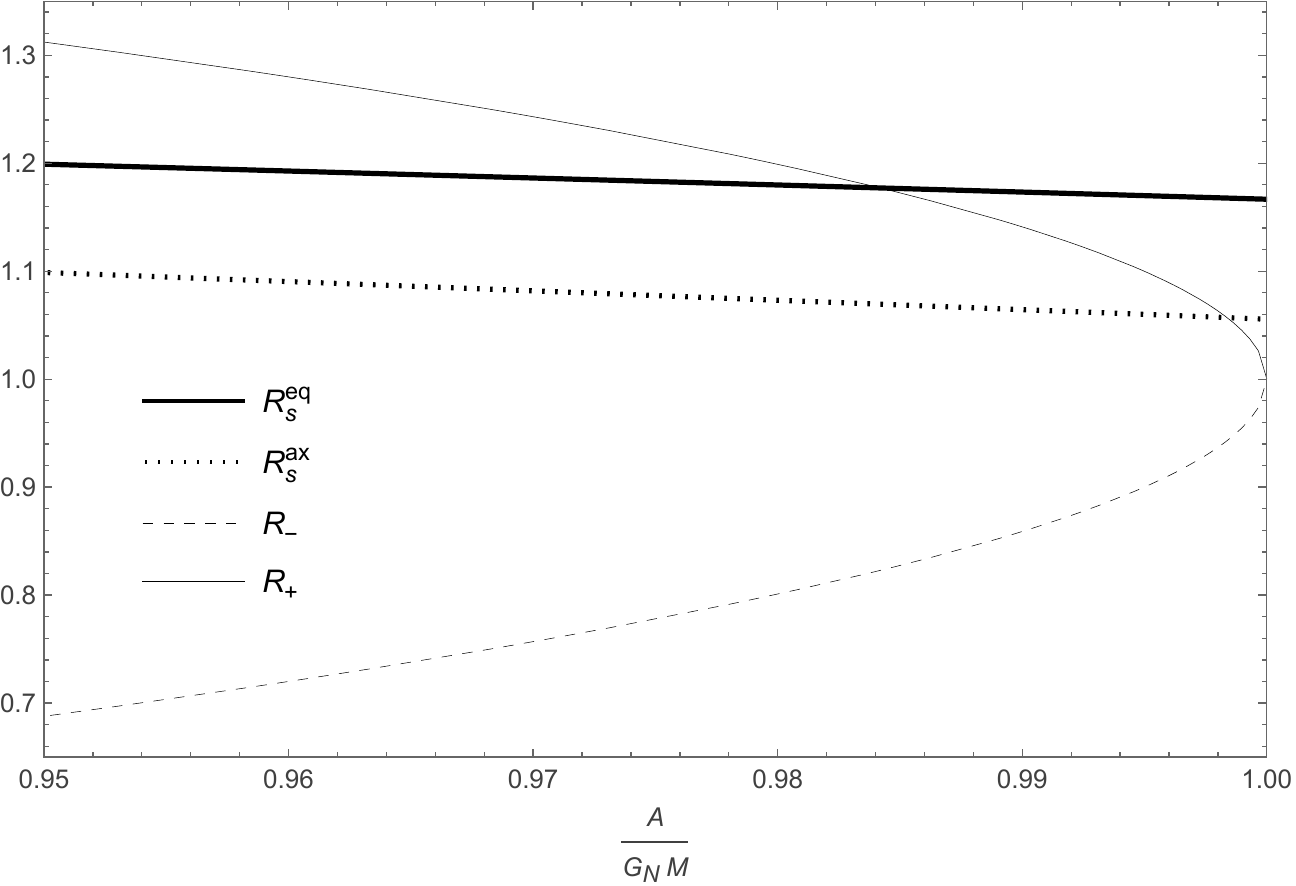}
\caption{Core radii {\rm vs\/} horizon radii for the whole range of classical Kerr black holes
$A^2\le \gn^2\,M^2$ (left panel); right panel shows the region of near extremal rotation magnified.}
\label{f:radii}
\end{figure}
\par
It is particularly interesting to compare the above results 
with the horizons~\eqref{R+-} of the (slowly rotating) external Kerr geometry,
\be
R_\pm
\simeq
\gn\,M
\pm
\gn\,M
\left(
1
-
\frac{A^2}{2\,\gn^2\,M^2}
\right)
\ .
\ee
From Fig.~\ref{f:radii}, we see that both core radii are larger than $R_-$ for all $A^2\le \gn^2\,M^2$.
The equatorial core radius remains shorter than $R_+$ for small rotation and until
$R_{\rm s}^{\rm eq}\simeq R_+$ for
\be
\frac{A^2}{\gn^2\,M^2}
\simeq
\frac{3}{2}\left(\sqrt{7}-2\right)
\simeq
0.984
\ ,
\ee
whereas the axial radius $R_{\rm s}^{\rm ax}\simeq R_+$ for
\be
\frac{A^2}{\gn^2\,M^2}
\simeq
\frac{3}{2}\left(\sqrt{73}-5\right)
\simeq
0.998
\ .
\ee
Clearly, both of these values of the specific angular momentum are way outside the
regime of slow rotation and should not be trusted.
\subsubsection{Near extremal rotation}
Let us consider the near extremal case for which $A_i^2\simeq\gn^2\,M_i^2$.
For the axial motion, we find
\be
W_i^{\rm ax}
\simeq
\frac{\gn\,M_i\,A_i^2}{R_i\left(R_i^2+A_i^2\right)}
\simeq
\frac{8\,A_i^2}{3\left(9\,\gn^2\,M_i^2+4\,A_i^2\right)}
\simeq
\frac{8}{39}
\ ,
\ee
whereas for equatorial motion
\be
W_i^{\rm eq}
\simeq
\frac{A_i^2}{2\,R_i^2}
\simeq
\frac{2}{9}
\ .
\ee
As noted before, these values are smaller than $V_i$ in Eq.~\eqref{V_i} so that previous perturbative
results should still provide at least a qualitatively valid picture in the whole classical range.
\subsection{Quantum rotating core}
\label{S:regular}
In the classical description, each dust layer can be made arbitrarily thin.
However, in the quantum description described above, we can assume the minimum thickness
is given by the uncertainty $\overline{\Delta R}_{n_i}\sim \bar R_{n_i}/3$ obtained from 
Eq.~\eqref{DR/R} for $n_i\gg 1$~\cite{Casadio:2023ymt}.
In the global ground state formed by layers in their own ground state we then have
\be
\bar R_{N_{i+1}}
=
\bar R_{N_{i+1}}^{(0)}
\left(1-2\,W_{i+1}\right)
=
\bar R_{N_{i}}
+
\overline{\Delta R}_{n_i}
\simeq
\frac{4}{3}\,
\bar R_{N_{i}}^{(0)}
\left(1-2\,W_{i}\right)
\ .
\ee
From Eq.~\eqref{bRNi}, we obtain
\be
\frac{3\,\lp\,M_{i+1}}{2\,\mpl}
\left(1-2\,W_{i+1}\right)
\simeq
\frac{2\,\lp\,M_{i}}{\mpl}
\left(1-2\,W_{i}\right)
\ ,
\ee
and the (discrete) mass function $M_i$ therefore depends on the angular momentum $A_i$.
In particular, on the symmetry axis, we have
\be
M_{i+1}^{\rm ax}
\left(1-\frac{16\,A_{i+1}^2}{27\,\gn^2\,(M_{i+1}^{\rm ax})^2}\right)
\simeq
\frac{4}{3}\,M_{i}^{\rm ax}
\left(1-\frac{16\,A_{i}^2}{27\,\gn^2\,(M_{i}^{\rm ax})^2}\right)
\ ,
\label{Mi1Miax}
\ee
whereas
\be
M_{i+1}^{\rm eq}
\left(1-\frac{4\,A_{i+1}^2}{9\,\gn^2\,(M_{i+1}^{\rm eq})^2}\right)
\simeq
\frac{4}{3}\,M_{i}^{\rm eq}
\left(1-\frac{4\,A_{i}^2}{9\,\gn^2\,(M_{i}^{\rm eq})^2}\right)
\ ,
\label{Mi1Mieq}
\ee
on the equator.
\par
From the general discussion in Ref.~\cite{Casadio:2023iqt}, we know that the inner Cauchy
horizon is not present if the mass function and specific angular momentum $m\sim a\sim r$.
Let us therefore assume that the specific angular momentum $A_i$ is linearly dependent on the
radius $R_i$,
\be
\label{angLin}
{A}_i 
\propto
\bar{R}_i
\simeq
\alpha\, \bar{R}_i^{(0)}
\ .
\ee
From Eqs.~\eqref{eq:axiPer} and~\eqref{eq:equPer}, the perturbations become
\be
W^{\rm{ax}}_i
\simeq
\frac{2\,\alpha^2}{3\,(1+\alpha^2)}
\simeq
\frac{2}{3}\,\alpha^2
\label{wAxL}
\ee
and
\be
W^{\rm{eq}}_i
\simeq
\frac{1}{2}\,\alpha^2
\ ,
\label{wEqL}
\ee
at leading order in the constant $\alpha$, which is of the same order of magnitude of the slow-rotation
parameter $A_i/\gn\,M_i$.
With this assumption, both Eqs.~\eqref{Mi1Miax} and \eqref{Mi1Mieq} therefore simplify to
\be
M_{i+1}^{\rm ax/eq}
\simeq
\frac{4}{3}\,M_{i}^{\rm ax/eq}
\ ,
\label{Mi1MiL}
\ee
which is precisely the linear behaviour found in the spherically symmetric case.
In fact, given the total ADM mass $M$, the mass distribution within each layer is then determined by
\be
M_{i}^{\rm ax/eq}
\simeq
\left( \frac{3}{4}  \right)^{N+1-i}
M
\equiv
M_i
\ ,
\ee
which implies that the quantum states with specific angular momentum given in
Eq.~\eqref{angLin} correspond to a mass function that grows linearly with the layer size, that is
\be
M_i
\simeq
\frac{2}{3} \left(1+\alpha^2  \right)
\frac{\bar{R}_{N_i}^{\rm eq}}{\gn} 
\simeq
\frac{2}{3} \left(1+\frac{4}{3}\,\alpha^2 
\right)
\frac{\bar{R}_{N_i}^{\rm ax}}{\gn} 
\ .
\label{mEq}
\ee 
We remark that this configuration is also consistent with the ellipsoidal shape of the collapsed cores
described by Eq.~\eqref{rAxEq}.
\par
For the above reasons, we will limit our following analysis to the case with linearly growing specific
angular momentum~\eqref{angLin} and mass function~\eqref{mEq}.
%
%
%
%
\subsection{Angular momentum and horizon quantisation}
\label{S:horQua}
According to Eq.~\eqref{NiW}, the ground state of dust particles in each layer is (approximately)
given by spherical wavefunctions~\eqref{groundI} with different quantum numbers along the axis
of symmetry and on the equatorial plane, namely
\be
N_i^{\rm eq}
\simeq
\frac{\mu\,M_i}{\mpl^2}
\left(1-\frac{1}{2}\,\alpha^2\right)
<
N_i^{\rm ax}
\simeq
\frac{\mu\,M_i}{\mpl^2}
\left(1-\frac{2}{3}\,{\alpha^2}\right)
\ .
\ee
Since $N_{i}^{\rm{ax}}$ and $N_{i}^{\rm{eq}}$ are integers, one obtains that the specific 
angular momentum must satisfy the quantisation rule
\be
N_{i}^{\rm{ax}}
-
N_{i}^{\rm{eq}}
\simeq
\left( \frac{3}{4}  \right)^{N+1-i}
\frac{\mu\,M}{6\,\mpl^2}
\,\alpha^2
\ .
\label{QuantAng}
\ee
For $i=N+1$, one obtains $M_i=M$ and
\be
\alpha^2
\simeq
\frac{6\,\mpl^2}{\mu\,M}
\left(
N_{N+1}^{\rm{ax}}
-
N_{N+1}^{\rm{eq}}
\right)
\ .
\ee
so that Eq.~\eqref{angLin} implies the quantisation of the specific angular momentum
in the exterior Kerr geometry
\be
A^2
\simeq
\left(\alpha\,R_{\rm s}^{(0)}\right)^2
\simeq
\frac{27\,M}{2\,\mu}
\,\lp^2
\left(
N_{N+1}^{\rm{ax}}
-
N_{N+1}^{\rm{eq}}
\right)
\ ,
\label{angQuant}
\ee
where we recall that $R_{\rm s}^{(0)}\simeq 3\,\gn\,M/2$.
\par
From Eq.~\eqref{RsW}, we also find
\be
\alpha^2
\simeq
\frac{R_{\rm s}^{\rm eq}-R_{\rm s}^{\rm ax}}{2\,\gn\,M}
\simeq
\frac{3\left(R_{\rm s}^{\rm eq}-R_{\rm s}^{\rm ax}\right)}{4\,R_{\rm s}^{(0)}}
\ ,
\ee
which implies 
\be
\frac{R_{\rm s}^{\rm eq}-R_{\rm s}^{\rm ax}}{R_{\rm s}^{(0)}}
\simeq
\frac{24\,\mpl^4}{\mu^2\,M^2}
\left(
N_{N+1}^{\rm{ax}}
-
N_{N+1}^{\rm{eq}}
\right)
\ ,
\ee
so that the shape of the core is also quantised.
\par
The outer horizon area in the Kerr geometry is given by
\be
\mathcal{A}_{\rm{H}}
=
4\, \pi 
\left(R_+^2+A^2\right)
\simeq
4\,\pi
\left(
R_{\rm s}^{(0)}
\right)^2
\left(
1
-
\frac{3}{4}\,\alpha^2
\right)
\ .
\ee
From the quantisation of the spherical dust core~\cite{Casadio:2023ymt},
\be
\frac{M}{\mu}\,N_{N+1}^{(0)}
\simeq
\frac{M^2}{\mpl^2}
\ ,
\ee
we then find
\be
\mathcal{A}_{\rm{H}}
\simeq
4\,\pi\,\lp^2\,
\frac{M}{\mu}\,
N_{N+1}^{(0)}
\left[
1
-
\frac{9\,\mpl^2}{2\,\mu\,M}
\left(
N_{N+1}^{\rm{ax}}
-
N_{N+1}^{\rm{eq}}
\right)
\right]
\ ,
\ee
where we remark that $M/\mu$ is not necessarily an integer.
\subsection{Effective interior geometry}
\label{s:effGeo}
We have seen that, for a core characterised by the linear relation~\eqref{angLin} for
the specific angular momentum, the mass function
$M_{i}^{\rm ax}\simeq M_{i}^{\rm eq}\equiv M_i\propto \bar R_i^{(0)}$.
It then follows that the corresponding effective metric can be simply obtained by applying the 
Gurses and Gursey algorithm~\cite{Gurses:1975vu} (generalised in Ref.~\cite{Casadio:2023iqt})
with a seed geometry provided by the spherically symmetric case of non-rotating dust described
in Refs.~\cite{Casadio:2023ymt,Bambagiotti:2025qxj} and specific angular momentum~\eqref{angLin}. 
This is all consistent with the initial assumption of a metric of the form in Eq.~\eqref{kerr},
where $m=m(r)$ and $a=a(r)$ are now determined by the ground states of the layers.
\par
The explicit form of the metric function $\Delta = \Delta(r)$ in Eq.~\eqref{e:delta}
for the ground state is particularly interesting since it allows us to verify that a Cauchy horizon
never appears. 
For this analysis, we will consider two effective mass functions, $m=m_{\rm par}(r)$
and $m=m_{\rm int}(r)$, which interpolate between the
interior discrete mass distribution $M_i$ and the outer Kerr metric with constant $M$
(for more details see also Ref.~\cite{Bambagiotti:2025qxj}).
In particular, 
\be
m_{\rm int}
=
c_{\rm i}\, r
\ , 
\quad
\text{for} \; 
0\le r<\bar{R}_{N_N}^{(0)}
\ ,
\ee
where the constant
\be
c_{\rm i}
=
\frac{M_N}{\bar{R}_{N_N}^{(0)}}
\simeq
\frac{2\,\mpl}{3\,\lp}
\ ,
\ee
and the analytic expression inside the outermost layer $\bar{R}_{N_N}^{(0)}<r<R_{\rm s}$ 
is given in Eq.~\eqref{B2} of Appendix~\ref{a:mass}.
\par
The function $m_{\rm par}$ is a parabolic profile which matches the outer constant ADM mass $M$ 
for $r\to\infty$, and was introduced in Ref.~\cite{Gallerani:2025wjc} as an
improvement over $m_{\rm int}$ to better account for the spatial overlapping of the
wavefunctions~\eqref{groundI}.
Its analytic expression is recalled in Eq.~\eqref{m_par}, from which we have
\be
m_{\rm par}
=
c_{\rm p}\, r
+
\mathcal{O}(r^2)
\ ,
\quad
\text{for} \; r \to 0^+
\ ,
\ee
with 
\be
c_{\rm p}
=
\frac{\bar{a}\,\mpl}{2\,\lp}
\ee
and the fitting parameter $\bar{a}\simeq 1.53$. 
\par
The effective angular momentum $a=a(r)$ in Eq.~\eqref{kerr} can be obtained from the above
mass functions by noting that Eqs.~\eqref{angLin} and \eqref{mEq} imply
\be
\frac{A_i}{A}
\simeq
\frac{\bar{R}_{N_i}^{(0)}}{R_s^{(0)}}
\simeq
\frac{M_i}{M}
\ ,
\label{effAng}
\ee
so that 
\be
a(r)
=
A\,\frac{m(r)}{M}
\ .
\ee
\par
\begin{figure}[t]
\centering
\includegraphics[width=0.48\textwidth]{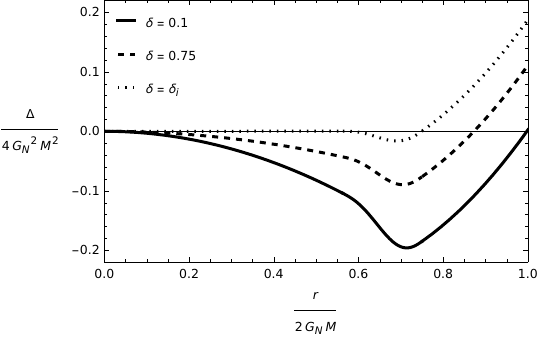}
$\quad$
\includegraphics[width=0.48\textwidth]{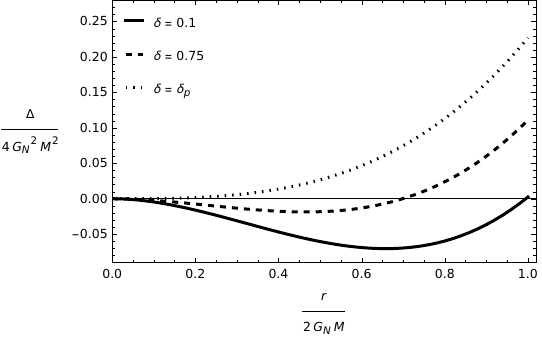}
\caption{Function $\Delta$ in Eq.~\eqref{e:delta} for $m_{\rm int}$ (left panel) and 
$m_{\rm par}$ (right panel) for different values of $\delta=A/\gn\, M$, with
$M=150 \, \mpl$.
The dotted lines correspond to the maximum values $\delta_{\rm i} \simeq 0.867$ (left panel)
and $\delta_{\rm p} \simeq 0.953$ (right panel).}
\label{f:delta}
\end{figure}

The function $\Delta = \Delta(r)$ is shown in Fig.~\ref{f:delta}.
In agreement with the general analysis of Ref.~\cite{Casadio:2023iqt},
we see that $\Delta(0)=0$ and there is a unique zero $\rh >0$ for values of the specific angular momentum
within the range
\be
0
<
\frac{A}{\gn\, M}
<
\delta_{\rm i} \equiv \frac{3}{2\,\sqrt{3}}
< 1
\ ,
\label{e:intRange}
\ee
for $m=m_{\rm int}(r)$, and
\be
0
<
\frac{A}{\gn\, M}
<
\delta_{\rm p}
\equiv
2\,\sqrt{\frac{\bar{a} -1}{\bar{a}^2}}
< 1
\ ,
\label{e:parRange}
\ee
for $m=m_{\rm par}(r)$.
For the critical values $\delta=\delta_{\rm i}$ or $\delta=\delta_{\rm p}$ the radius $\rh=0$. 
\par
Note that both ranges in Eqs.~\eqref{e:intRange} and~\eqref{e:parRange} impose stronger
requirements on $A/\gn M$ for the existence of the horizon than the limits of the perturbative regime
given in Eqs.~\eqref{e:axPer} and~\eqref{e:eqPer}.
In particular, this analysis for $m_{\rm int}$ and $m_{\rm par}$ excludes classical near-extremal
configurations with $A\simeq \gn M$.
\section{Conclusions and outlook}
\setcounter{equation}{0}
\label{S:conc}
In this work we extended the quantisation of the layered dust core~\cite{Casadio:2023ymt}
to rotating (ellipsoidal) cores, where dust particles move along time-like geodesics of the 
(generalised) Kerr geometry~\eqref{kerr}, improving the description given in Ref.~\cite{Casadio:2022epj}.
In that previous work, particles fall freely along geodesics in the Schwarzschild spacetime,
and their rotation was only approximately described by states with non-zero angular momentum,
hence neglecting the general-relativistic features of the Kerr geometry. 
\par
If we only consider motion along the rotation axis or on the equator,
the Hamiltonian constraint~\eqref{SchE} reduces to the expression 
for the spherically symmetric distribution, described in detail in 
\cite{Casadio:2021cbv , Casadio:2023ymt}, with an additional term 
$W_i \sim a^2/\gn^2\, M^2$ which can be treated as a perturbation.
The global ground state is described by ellipsoidal layers elongated on the equator
with respect to the rotation axis, that form a core with size smaller than the spherically symmetric
dust distribution.
In the perturbative regime of slow-rotation, the equatorial radius
$\bar{R}_{\text{s}}^{\text{eq}}$ is always shorter than the outer horizon $R_+$,
thus describing a black hole, whereas a configuration with $R_{\text{s}}^{\text{ax/eq}}\simeq R_+$
cannot be described within this perturbative approach.
We remark that the discrepancy with the results in Ref.~\cite{Casadio:2022epj}
is due to neglecting the general-relativistic effects of the Kerr solution,
in which the angular momentum introduces both an attractive and repulsive term in the
radial geodesic motion.
\par
A possible ground state appears to be described by a configuration in which $A_i \propto \bar{R_i} \propto M_i$, 
as described in Section~\ref{S:regular}. 
In fact, this condition ensures that no Cauchy horizon ever forms within
the core and that the central ring-singularity is replaced by an integrable singularity.
In this configuration, the specific angular momentum $A_i$ is quantized according to
Eq.~\eqref{QuantAng}, which ensures the quantisation of the specific angular momentum $A$
in the exterior solution, in agreement with Ref.~\cite{Casadio:2022epj}.
Together with the quantisation of the spherical dust core, this result
implies that the outer horizon area is quantised in Plank units.
Remarkably, in this configuration the effective metric can be determined from the spherically symmetric
non-rotating dust distribution given in Refs.~\cite{Casadio:2021cbv, Casadio:2023ymt,Gallerani:2025wjc}. 
In Section~\ref{s:effGeo}, we analysed the effective causal structure inside the core by means of
the function $\Delta=\Delta(r)$, which is computed for the specific seed mass distributions described
in Ref.~\cite{Bambagiotti:2025qxj}.
The absence of a Cauchy horizon and the existence of the only event horizon at $r=\rh$, within the
slow-rotation limits given by Eqs.~\eqref{e:intRange} and~\eqref{e:parRange}, is in agreement with
the general picture that quantum matter should be able to regularise the interior region of black
holes~\cite{Casadio:2023iqt}.
\par
We have not considered possible observational signatures in the present work.
First of all, the interior core geometry remains classically unaccessible to observations
due to the outer horizon.
Moreover, it is not clear if any classical observables can be extracted from the interior
effective geometry, as we already discussed in Section~\ref{S:intro}.
For example, classical geodesic (or other) motion inside the core should not occur
and only the response of the whole core to outer perturbations during disruptive processes,
like the merging of a binary black hole system, could be determined from this quantum state.
On the other hand, since the geometry outside the core is given by the classical Kerr metric,
we expect deviations from General Relativity be qualitatively similar to those found in
Ref.~\cite{Bambagiotti:2025qxj}
for the spherically symmetric case and mainly depend on the mass function around the
core surface (see also Appendix~\ref{a:mass}).
In particular, the spectrum of quasi-normal modes should be sensitive to any deviations
from the outer Kerr geometry provided the core has a size comparable to the radius of the
event horizon that will further depend on the detailed quantum description of the core surface.
Such developments are left for future investigations.
\section*{Acknowledgments}
T.B.~and R.C.~are partially supported by the INFN grant FLAG. 
The work of R.C.~has also been carried out in the framework of activities of the National Group of
Mathematical Physics (GNFM, INdAM) and the COST action CA23115 (RQI).
\appendix
\section{Effective mass functions}
\label{a:mass}
We show here the explicit forms of the continuous MSH mass functions used in Section~\ref{s:effGeo}.
More details can be found in Ref.~\cite{Bambagiotti:2025qxj}.
\par
The mass function $m=m_{\rm int}(r)$ simply interpolates between the linear behaviour $M_i\propto \bar R_i$
up to the inner radius of the outermost layer and the constant ADM mass $M$ outside the core,
\be
m_{\rm int}
=
\begin{cases}
c_{\rm i} \, r 
\ ,
\qquad 
&{\rm for}
\quad
r \le \bar{R}_{N_N}^{(0)}
\\ 
B(r)
\ ,
\qquad 
&{\rm for}
\quad
\bar{R}_{N_N}^{(0)} \leq r \leq R_s^{(0)}=\frac{4}{3}\,\bar{R}_{N_N}^{(0)}
\\
M
\ ,
\qquad 
&{\rm for}
\quad
r \geq R_s^{(0)}
\ .
\end{cases}.
\ee
The interpolating function is given by the 5-th order polynomial
\begin{align}
B
=
&
-\frac{c_1}{\Delta x^3}
\left(x-x_1\right)^3
+\frac{M}{\Delta x^3}
\left(x-x_0\right)^3 
-\left( \frac{3\, c_1}{\Delta x^4} + \frac{c_2}{\Delta x^3} \right)
\left(x-x_1\right)^3
\left(x-x_0\right)
\notag
\\
&
-\frac{3\,M}{\Delta x^4}
\left(x-x_0\right)^3
\left(x-x_1\right)
-\left( \frac{6\,c_1}{\Delta x^5} + \frac{3\,c_2}{\Delta x^4} \right)
\left(x-x_1\right)^3
\left(x-x_0\right)^2
\notag
\\
&
+\frac{6\,M}{\Delta x^5}
\left(x-x_1\right)^2
\left(x-x_0\right)^3
\ ,
\label{B2}
\end{align}
where $x\equiv r/2\,\gn\,M$, $c_1= M_N$, $c_2= c_1\,(2\,\gn\,M/\bar{R}_{N_N}^{(0)})$,
and the width of the outermost layer is denoted by $\Delta x = (R_{\rm s}^{(0)}-\bar R_{N_N}^{(0)})/2\,\gn\,M=1/4$.
\par
The wavefunctions of dust particles~\eqref{groundI} do not vanishing outside the respective layers,
which yields a non-zero probability that at least the nearest layers overlap.
To account for this effect, the mass function $m=m_{\rm par}(r)$
was computed in Ref.~\cite{Gallerani:2025wjc} for the spherically symmetric case and is given by
\be
m_{\rm par}
=
M \left( \bar{a}\,x+\bar{b}\,x^{\bar{c}} \right)
\ ,
\label{m_par}
\ee
with $x\equiv r/2\,\gn\,M$ and the fitting coefficients
\be
\bar{a}=1.53\ ,
\quad
\bar{b} = -0.533
\ ,
\quad
\bar{c} = 1.90
\ .
\ee
These values were used for the plots in the right panel of Fig.~\ref{f:delta}.
\end{document}